\documentclass[]{aastex631}

\begin{document}

\title{Three-dimensional velocity fields of the solar filament eruptions detected by CHASE}

\correspondingauthor{Chuan Li}
\email{lic@nju.edu.cn}

\correspondingauthor{Zhen Li}
\email{lizhen@nju.edu.cn}

\author[0000-0002-1190-0173]{Ye Qiu}
\affiliation{Institute of Science and Technology for Deep Space Exploration, Suzhou Campus, Nanjing University, Suzhou 215163, China}

\author[0000-0001-7693-4908]{Chuan Li}
\affiliation{Institute of Science and Technology for Deep Space Exploration, Suzhou Campus, Nanjing University, Suzhou 215163, China}
\affiliation{School of Astronomy and Space Science, Nanjing University, Nanjing 210023, China }
\affiliation{ Key Laboratory for Modern Astronomy and Astrophysics (Nanjing University), Ministry of Education, Nanjing 210023, China}

\author[0000-0002-9293-8439]{Yang Guo}
\affiliation{School of Astronomy and Space Science, Nanjing University, Nanjing 210023, China }
\affiliation{ Key Laboratory for Modern Astronomy and Astrophysics (Nanjing University), Ministry of Education, Nanjing 210023, China}

\author{Zhen Li}
\affiliation{School of Astronomy and Space Science, Nanjing University, Nanjing 210023, China }
\affiliation{ Key Laboratory for Modern Astronomy and Astrophysics (Nanjing University), Ministry of Education, Nanjing 210023, China}

\author[0000-0002-4978-4972]{Mingde Ding}
\affiliation{School of Astronomy and Space Science, Nanjing University, Nanjing 210023, China }
\affiliation{ Key Laboratory for Modern Astronomy and Astrophysics (Nanjing University), Ministry of Education, Nanjing 210023, China}

\author[0000-0003-1350-9722]{Linggao Kong}
\affiliation{Institute of Science and Technology for Deep Space Exploration, Suzhou Campus, Nanjing University, Suzhou 215163, China}

\begin{abstract}
The eruption of solar filaments, also known as prominences appearing off-limb, is a common phenomenon in the solar atmosphere. It ejects massive plasma and high-energy particles into interplanetary space, disturbing the solar-terrestrial environment. It is vital to obtain the three-dimensional velocity fields of erupting filaments for space-weather predictions. We derive the three-dimensional kinematics of an off-limb prominence and an on-disk filament, respectively, using the full-disk spectral and imaging data detected by the Chinese H$\alpha$ Solar Explorer (CHASE). It is found that both the prominence and the filament experience a fast semicircle-shaped expansion at first. The prominence keeps propagating outward with an increasing velocity until escaping successfully, whereas the south leg of the prominence finally moves back to the Sun in a swirling manner. For the filament, the internal plasma falls back to the Sun associated with an anticlockwise rotation in the late ejection, matching the failed eruption without a coronal mass ejection.  During the eruptions, both the prominence and the filament show material splitting along the line-of-sight direction, revealed by the bimodal H$\alpha$ spectral profiles. For the prominence, the splitting begins at the top and gradually spreads to almost the whole prominence with a fast blue-shift component and a slow red-shift component. The material splitting in the filament is more fragmental. As shown by the present results, the CHASE full-disk spectroscopic observations make it possible to systematically study the three-dimensional kinematics of solar filament eruptions. 

\end{abstract}

\keywords{Solar filaments (1495) --- Solar filament eruptions (1981) --- Solar coronal mass ejections (310) --- Spectroscopy (1558)}

\section{Introduction} \label{sec:intro}
Due to the lower-temperature and higher-density characteristics compared to the surrounding coronal plasma, solar filaments appear dark on the disk, whereas their off-limb counterparts -- prominences look bright against the dark space background~\citep{2021ApJ...917...81G, 2022NatAs...6..942J}. Some of them can be triggered to erupt by magnetic reconnection \citep{1999ApJ...510..485A,2000ApJ...545..524C,2001ApJ...552..833M,2021NatAs...5.1126J}, magnetohydrodynamic instabilities \citep{2004A&A...413L..27T,2006PhRvL..96y5002K,2021ApJ...908...41A}, or both \citep{2023A&A...677A..43P}. The eruptive filaments escaping from the Sun evolve into bright cores of coronal mass ejections (CMEs; \citealt{1981ApJ...244L.117H}), which presumably induce geomagnetic storms and pose threats to terrestrial high-technology systems \citep{1982Natur.299..537M}. Thus, it is imperative to acquire the dynamics of eruptive filaments to figure out their physical mechanisms and predict the latent geomagnetic storms caused by the eruptions.

A general method to get the kinematics of ascending filaments is making a slice to track its plane-of-sky (POS) apex and further fitting the trajectory with an exponential-plus-polynomial function \citep{2020ApJ...889...28M,2020ApJ...901...13Q}. It is found that the filament eruptions usually comprise a preceding slow-rise phase and a following fast-rise phase. The turning point between the two phases is defined as the point when the velocity fitted by the polynomial component is identical to the one by the exponential component \citep{2015SoPh..290.1703M,2020ApJ...894...85C}. However, this method is unable to measure the kinematic parameters along the line-of-sight (LOS) direction. One solution is the three-dimensional geometry reconstruction based on the quasi-simultaneous observations from at least two different perspectives \citep{2011ApJ...730..104J,2011ApJ...739...43L,2020ApJ...897...35R}. It, however, requires an appropriate separation angle between the Earth-orbit satellite and the off-Sun-Earth-line missions, e.g., Solar TErrestrial RElations Observatory (STEREO; \citealt{2008SSRv..136....5K}), Parker Solar Probe (PSP; \citealt{2016SSRv..204....7F}), and Solar Orbiter (SolO; \citealt{2020A&A...642A...1M}). 

Another way to obtain the LOS velocities is to calculate the Doppler shift through spectroscopic observations (e.g., \citealt{2010A&A...514A..68S,2015ApJ...806....9K,2017ApJ...843L..24S}). There have been several studies verifying the feasibility of utilizing spectroscopic data to acquire the three-dimensional kinematics of solar eruptions \citep{2003PASJ...55..503M,2015ApJ...803...85L,2017ApJ...836...33C,2017A&A...606A..30S,2019A&A...624A..72Z,2021EP&S...73...58S,2021PASJ...73..394G,2023Ap&SS.368...40J}. Among solar spectral lines, H$\alpha$ is a good diagnostic line for solar filament observations. One way to detect the H$\alpha$ line is taking images of the Sun at several wavelengths from the line core to the wings using a tunable filter. This kind of instruments includes the Improved Solar
Observing Optical Network (ISOON; \citealt{1998ASPC..140..519N}), the Meudon Spectroheliograph \citep{2019SoPh..294...52M}, the Solar Dynamics Doppler Imager (SDDI; \citealt{2017SoPh..292...63I}) and the Flare Monitoring Telescope (FMT; \citealt{2007BASI...35..697U}). The disadvantage of such instruments is their low spectral resolution. The other way is observing the H$\alpha$ spectra through the grating spectrograph. In this type, the Fast Imaging Solar Spectrograph (FISS; \citealt{2013SoPh..288....1C}) at Big Bear Solar Observatory has very high spatial resolution of 0.2$''$ and spectral resolution of 0.019~\AA. It, however, has a limited field of view (FOV), thus cannot track the eruption of a large filament. 

The Chinese H$\alpha$ Solar Explorer (CHASE; \citealt{2022SCPMA..6589602L}) acquires, for the first time, the seeing-free H$\alpha$ spectroscopic observations of the full solar disk. It allows us to systematically study the three-dimensional kinematics of solar filament eruptions. In this Letter, we present a detailed investigation of the three-dimensional velocity fields of an off-limb prominence and an on-disk filament, respectively. A brief overview of the data and the two eruptions is given in Section \ref{sec:data}. Detailed methods of velocity derivation and results are shown in Section \ref{sec:res}. Section \ref{sec:dis_con} is devoted to discussion and conclusion.

\section{Observations} \label{sec:data}
The CHASE mission, launched on 2021 October 14 was designed to obtain the full-disk spectra at both H$\alpha$ (6562.8~\AA, a chromospheric spectral line) and Fe I (6569.2~\AA, a photospheric spectral line) passbands \citep{2022SCPMA..6589602L,2022SCPMA..6589603Q}. The wavelength in H$\alpha$ passband ranges from 6559.7~\AA~to 6565.9~\AA~corresponding to -141.7 km s$^{-1}$ and 141.7 km s$^{-1}$. The spatial and spectral resolutions of the equipped H$\alpha$ Imaging Spectrograph (HIS; \citealt{2022SCPMA..6589605L}) are 1.2$''$ and 0.072~\AA, respectively.  For both events, the observations are in binning mode with the spatial and spectral samplings of 1.04$''$ and 0.048~\AA, respectively. The temporal resolution is about 73 s and the FOV is 40$'$ $\times$ 40$'$, sufficient for tracing the whole eruptions of most ejected prominences. Moreover, we employ the extreme-ultraviolet (EUV) images obtained from the Atmospheric Imaging Assembly (AIA; \citealt{2012SoPh..275...17L}) on board the Solar Dynamics Observatory (SDO; \citealt{2012SoPh..275....3P}) to show the high-temperature counterparts of the filament eruptions. For the on-disk filament, we additionally use the photospheric magnetogram taken by the Helioseismic and Magnetic Imager (HMI; \citealt{2012SoPh..275..207S}) on board SDO.

\begin{figure}[ht!]
	\plotone{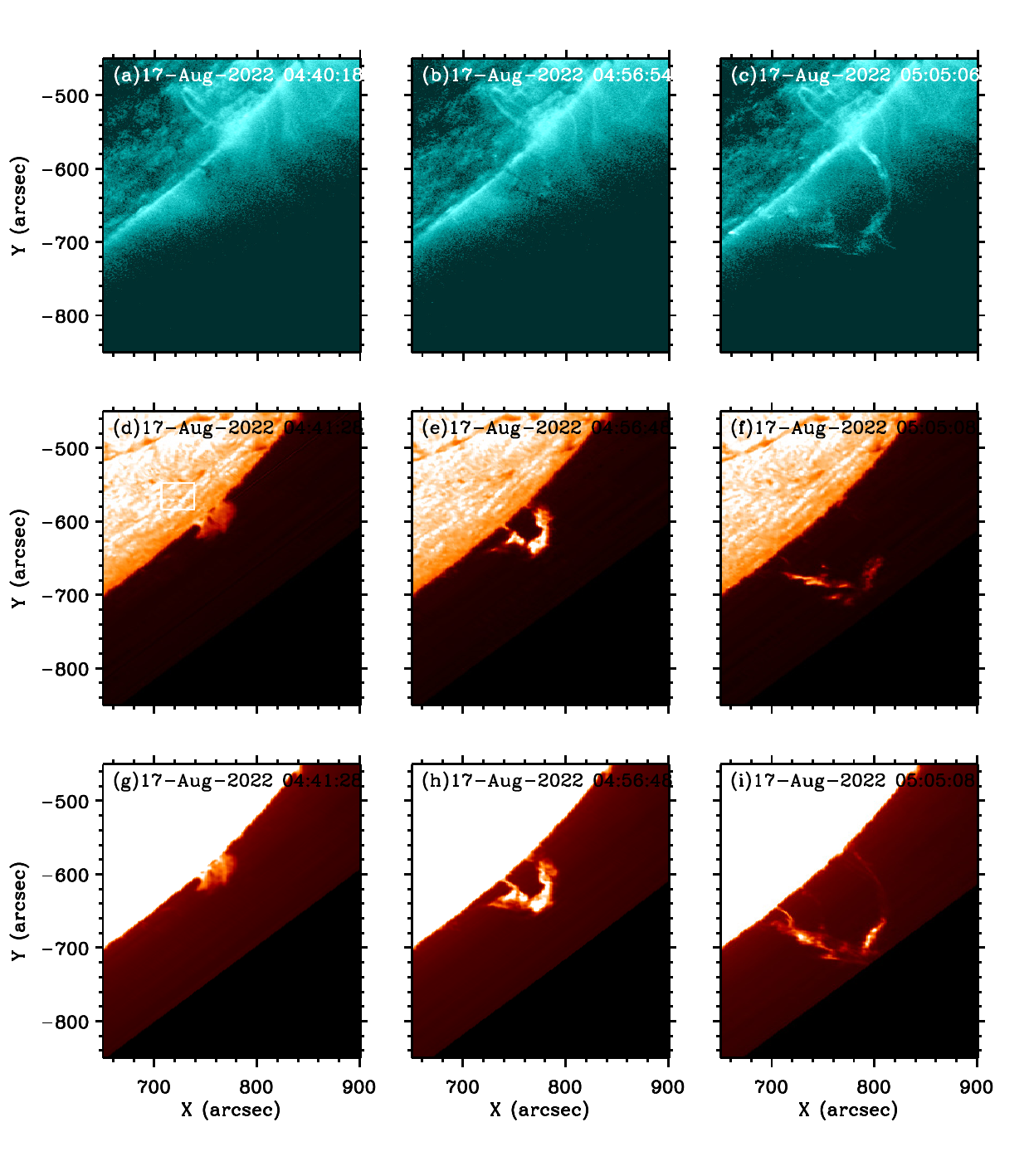}
	\caption{Overview of the eruptive prominence on 2022 August 17. (a)--(c)~\textit{SDO}/AIA 131~\AA~images. (d)--(f) CHASE H$\alpha$ line center images. The white rectangle marks the background region for zero Doppler shift. (g)--(i) Enhanced images corresponding to (d)--(f), which are weight-averaged along the wavelength dimension with a Gaussian function.} \label{fig:prominence1}
\end{figure}

The prominence eruption occurred at the southwest solar limb on 2022 August 17. It was associated with a CME captured by the Large-Angle Spectroscopic Coronagraph (LASCO) on board the Solar and Heliospheric Observatory (SOHO) at 05:12:05 UT. No flare was recorded probably due to the limb occultation. The detection of this event by CHASE started from 04:40:14 UT to 05:06:16 UT. As revealed by Figure~\ref{fig:prominence1}, the prominence presents an overall expansion at H$\alpha$ line center images as it ascends. At the beginning, the eruptive structure is too weak to be distinguished in the AIA 131~\AA~images. As the eruption goes on, the magnetic reconnection heats the outer edge of the prominence and brightens it to make it visible as a hot channel at EUV passbands \citep{2006A&A...458..965C}.

\begin{figure}[ht!]
	\plotone{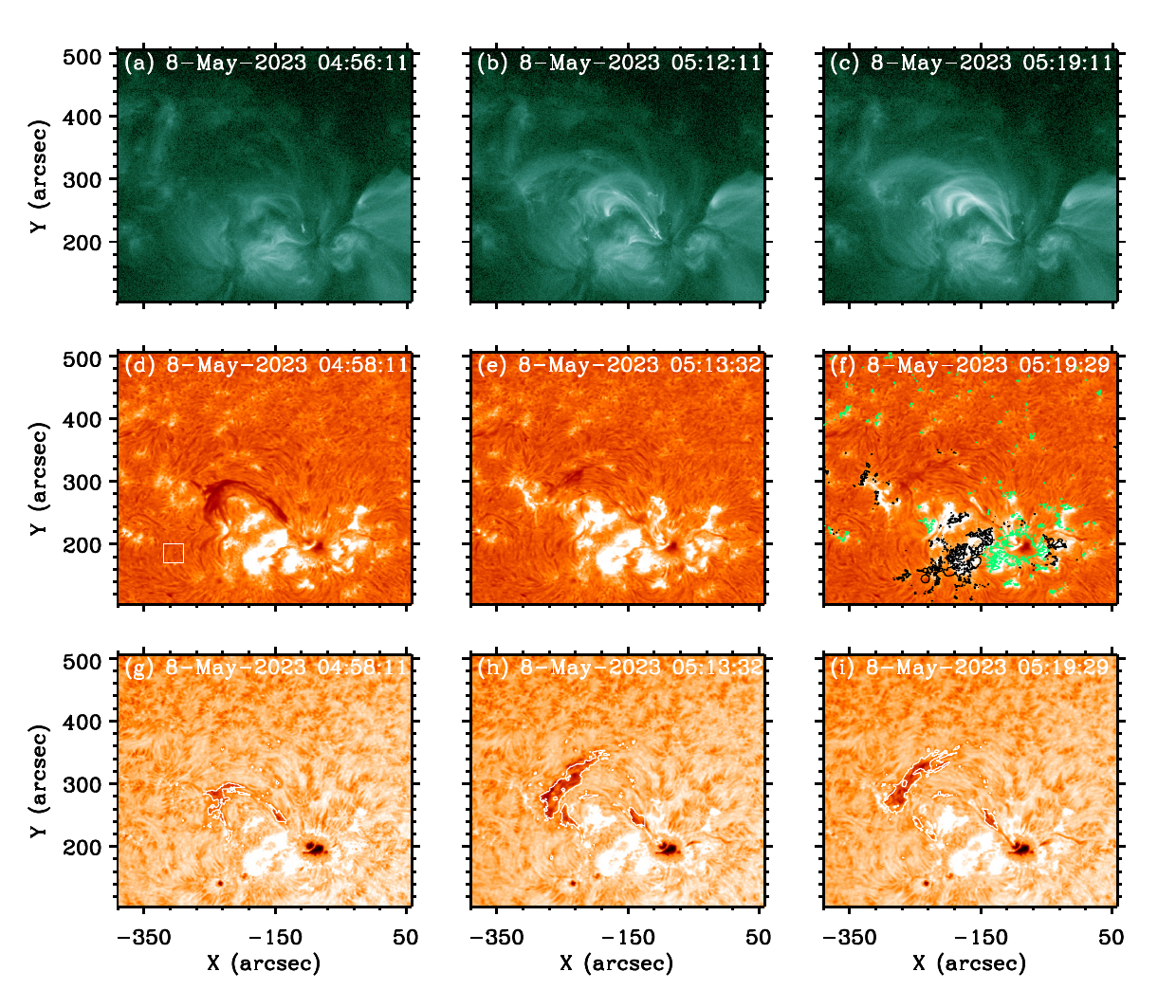}
	\caption{Overview of the eruptive filament on 2023 May 8. (a)--(c)~\textit{SDO}/AIA 94~\AA~images. (d)--(f) CHASE Ha line center images. The black and green contours mark the -300 G and 300 G LOS photospheric magnetic fields. The white rectangle in panel (d) marks the quiescent region chosen as the zero-shifted background. (g)--(i) Enhanced images corresponding to (d)--(f). The white curves outline the recognized filament region. \label{fig:filament1}}
\end{figure}

The filament eruption was located near the center of solar disk in the NOAA active region 13297 on 2023 May 8. It was accompanied by a C3.1 flare which started at 05:05 UT, peaked at 05:24 UT and ended at 05:34 UT. No CMEs were recorded, indicating a failed filament eruption. The observations involving this eruption of CHASE is accessible from 04:56:58 UT to 05:25:22 UT. Figures~\ref{fig:filament1}(a)--(c) show the processes from the flare onset to the formation of flare loops underneath the filament eruption at AIA 94~\AA~waveband. Figures 2(d)--(i) show the processes of filament eruption, disintegration, and drainage of filament materials. According to the drainage sites and the photospheric magnetic field distribution, the filament is determined to have a sinistral chirality \citep{2014ApJ...784...50C}.

\section{Methods and Results} \label{sec:res}

\subsection{Doppler velocity of the eruptive prominence}\label{subsec:doppler_pro}
The filament can be regarded as a lump of cold cloud-like plasma suspended in the hot corona, so the emission intensity of the filament can be calculated from the simplified radiative transfer equation in cloud model \citep{1964PhDT........83B,1988A&A...203..162M}: 
\begin{equation}\label{eq:original_cloud}
	I =  I_{0} e^{-\tau}+S\left(1-e^{-\tau}\right),
\end{equation}
where $I_0$ is the background intensity, $S$ is the source function of the filament assumed as a constant, and $\tau$ represents the optical depth dependent on the wavelength with a Gaussian form as below:
\begin{equation}
 \tau(\lambda)=\tau_{0} \exp \left[-\left(\lambda-\lambda_{0}-v \lambda_{0} / c\right)^{2} / W^{2}\right].
\end{equation}	
In this equation, $\tau_0$ is the value at the center of the Gaussian function, $\lambda_0$ is the line core of the H$\alpha$ profile averaged from the background region, $v$ is the derived Doppler velocity where negative values indicating towards the observer, and $W$ denotes the line width. The $S$, $\tau_0$, $v$ and $W$ are the parameters needing to be solved by fitting to the observed H$\alpha$ spectral profiles of the on-disk filament..

For the off-limb prominence, the background intensity is theoretically zero. We can treat the prominence as a luminous slab radiating with a constant source function: 
\begin{equation}\label{eq:one_pro}
	I =  N_0+S\left(1-e^{-\tau}\right),
\end{equation}
where the new parameter $N_0$ represents a value of instrumental effects, i.e., scattered light, which leads to a non-zero background emission.

During the ejection, some regions of the prominence may split into two pieces along the LOS direction, with one piece moving towards the observer and the other moving away, resulting in the bimodal spectral profiles as shown in Figure~\ref{fig:specfit} (c). In this case, we adopt the modified fitting function based on the two-cloud model \citep{2014ApJ...792...13H}:
\begin{equation}\label{eq:two_pro}
	I =  N_0+S_{\rm{b}}(1-e^{-\tau_{\rm{b}}})e^{-\tau_{\rm{f}}}+S_{\rm{f}}(1-e^{-\tau_{\rm{f}}}),
\end{equation}
where the subscript ``b" denotes the backward piece and ``f" represents the piece towards the observer.

We then fit the H$\alpha$ profiles of the studied prominence, as shown typically in Figures~\ref{fig:specfit}(a)--(c), using Equations (\ref{eq:one_pro}) and (\ref{eq:two_pro}). The uncertainty of the Doppler velocity caused by the fitting is 2.47 km s$^{-1}$, a mean value in the prominence region at every sequence, while the uncertainty introduced by the wavelength calibration is 1.12 km s$^{-1}$, leading to a total uncertainty of around 2.71 km s$^{-1}$. As shown in Figures~\ref{fig:prominence2}(a)--(f) (corresponding animation is available online), there exists slight material flow inside the prominence before its eruption. The prominence starts to rise at 04:46:12 UT. At the early stage of the eruption, the prominence shows an overall expansion with a north redshifted leg and a south blueshifted leg. The south leg falls back to the Sun due to the gravity at 04:58:02 UT, converting the south blueshifted leg into redshifted. The large-area splitting of the prominence begins at 04:52:04 UT. The backward piece exhibits weak blueshift and then redshift with the magnitude of 10 km s$^{-1}$. The forward piece is dominated by high blueshift of around 80 km s$^{-1}$, meaning this piece moving fast towards the Earth.

\subsection{POS velocity of the eruptive prominence}\label{subsec:transverse_pro}
We adopt Fourier Local Correlation Tracking (FLCT; see details in \citealt{2004ApJ...610.1148W,2007ApJ...670.1434W}) approach to derive the POS velocity of the prominence eruption. Before the computation, we need to enhance the prominence signal in the two-dimensional images by weighted-averaging each three-dimensional data along the wavelength dimension, because the erupting prominence is a multi-velocity structure that cannot be wholly displayed with images at a single wavelength. The enhanced images are shown in Figures~\ref{fig:prominence1}(g)--(i). The weighted average function is a Gaussian function centered on the background H$\alpha$ line core with a full width at half maximum of 1.5~\AA. Figures~\ref{fig:prominence2}(g)--(i) display the POS velocities of the erupting prominence. It is found that the prominence experiences an overall expansion at the beginning as mentioned above and the expansion goes faster and faster with the evolution of the eruption reaching the magnitude of 100 km s$^{-1}$. Based on the derived LOS and POS velocities, the maximum ejection velocity of the prominence is about 160 km s$^{-1}$ with 22$^\circ$ inclined to the POS. Besides, the drainage of the south leg falls back in a swirling way after 05:02:46 UT.

\subsection{Doppler velocity of the eruptive filament}\label{subsec:doppler_fil}
Before applying the cloud model to a filament, we first need to identify the filament region. Since the H$\alpha$ profile of a filament has a deeper line depth than the surroundings, we compare the integrated intensity between H$\alpha \pm$1.5~\AA~of the spectral lines, which have already been normalized to the continuum intensity. The data outside this wavelength range is not used because the absorption of the filament is not obvious in the far wings. The threshold is set to 0.9 times the integrated intensity in the same waveband of the normalized background line profile. All the locations with the integrated intensity less than the threshold are recognized as the filament regions , which are marked with white contours in Figures~\ref{fig:filament1}(g)--(i).

Equation~(\ref{eq:original_cloud}) considers only a single position in the filament. In fact, a filament can spread over a large area on the solar disk, thus the intensity along the filament is affected by the limb darkening effect. Taking the limb darkening into account, we introduce a new fitting parameter $C_{\rm ld}$. In a very narrow wavelength range, the change of limb darkening coefficient can be neglected, especially at the location near the disk center \citep{1998A&A...333..338H}. Therefore, we can treat $C_{\rm ld}$ at each spatial point as a constant independent on wavelength. Then the single-cloud function can be written as:
\begin{equation}\label{eq:one_fil}
  I =  C_{\rm ld} I_{0} e^{-\tau}+S\left(1-e^{-\tau}\right).
\end{equation}

\begin{figure}[ht!]
	\plotone{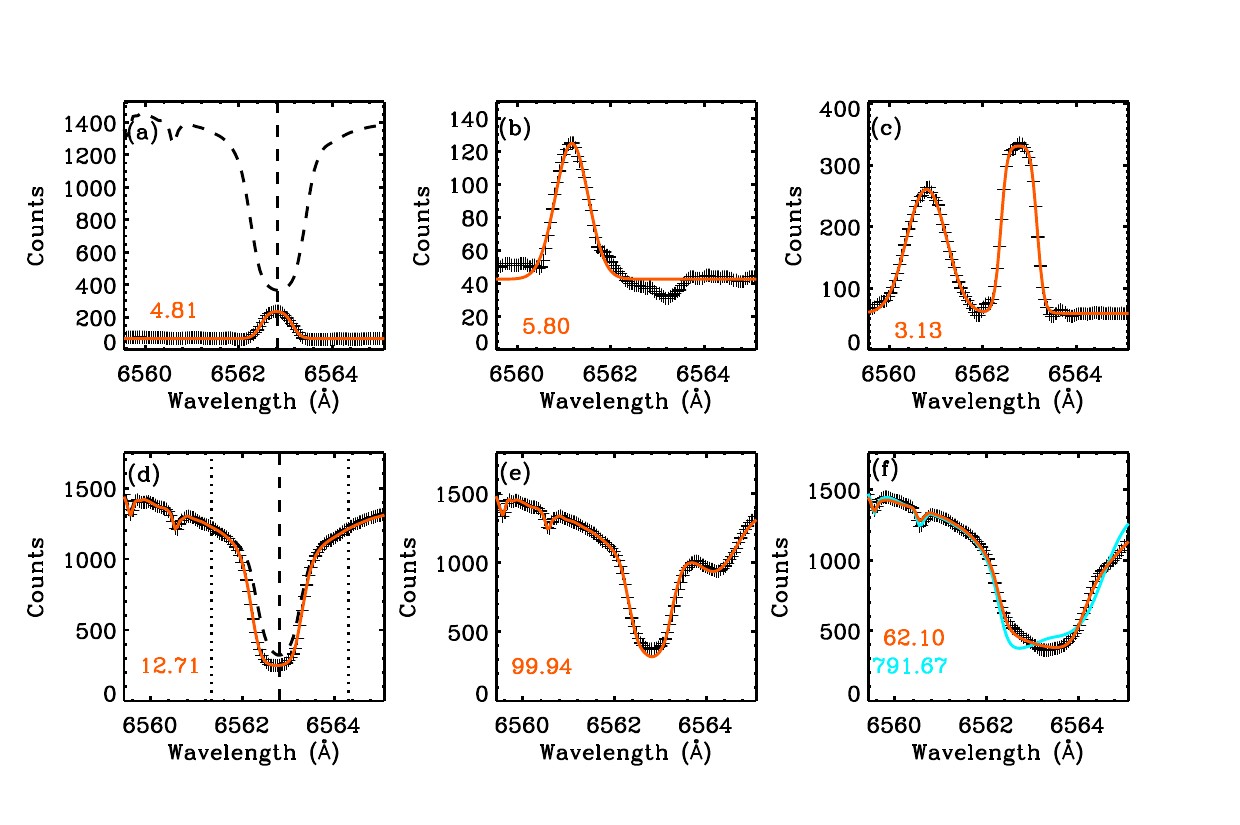}
	\caption{Spectral profiles and cloud-model fitting results of the prominence (a--c) and filament (d--f). The dashed curve is the averaged profile of  background region, with the dashed vertical line marking its line center. The plus points represent the observed profiles, while the colored curves exhibit the cloud model fitting results with the corresponding chi-square values given at lower left. The cyan curve in panel (f) is the single-cloud fitting result, and the orange is double-cloud.  \label{fig:specfit}}
\end{figure}

\begin{figure}[ht!]
	\plotone{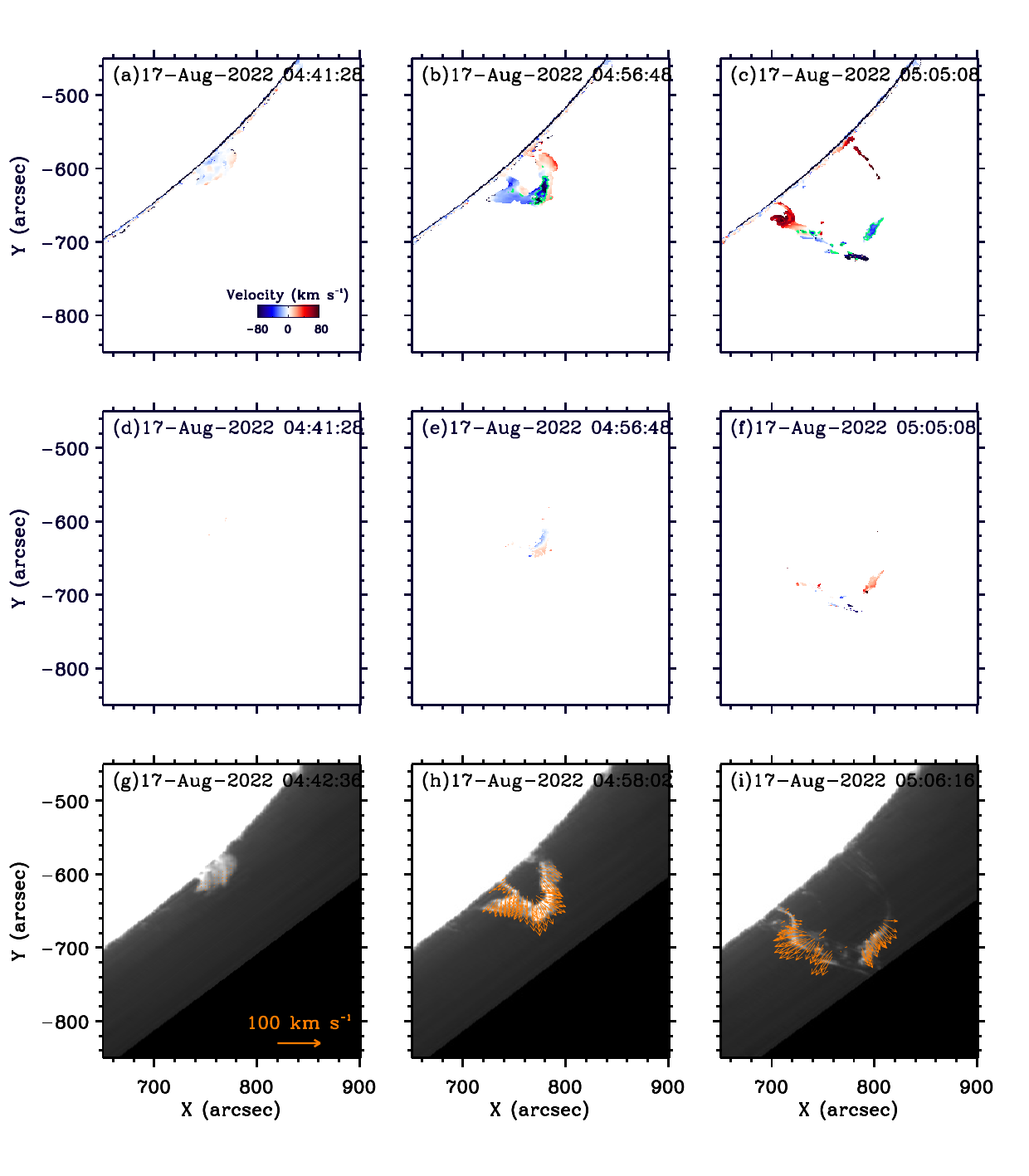}
	\caption{Three-dimensional velocity fields of the prominence. (a)--(c) Doppler velocity deduced by the single-cloud model and the front component of the double-cloud model. The green lines sketch the region of the double-cloud model fitting. (d)--(f) Doppler velocity of the back component of the double-cloud model. (g)--(i) POS velocity calculated by the FLCT method, where an orange arrow indicates the value and direction of the velocity in a local position. A 2-second animation is attached online, exhibiting the evolution of the three-dimensional velocities of the prominence from 04:41:28 UT to 05:05:16 UT. \label{fig:prominence2}}
\end{figure}

Different from the prominence, the bimodal structure of the filament in Figure~\ref{fig:specfit}(e) originates from the absorptions of both the background chromospheric plasma and the moving filament, which also can be well-fitted by the single-cloud model. However, there are some irregular H$\alpha$ line profiles unable to be well-fitted by the single-cloud function. For these profiles, we use the modified two-cloud mo\label{key}del:
\begin{equation}\label{eq:two_fil}
  I =  C_{\rm ld} I_{0} e^{-(\tau_{\rm{b}}+\tau_{\rm{f}})}+S_{\rm{b}}(1-e^{-\tau_{\rm{b}}})e^{-\tau_{\rm{f}}}+S_{\rm{f}}(1-e^{-\tau_{\rm{f}}}).
\end{equation}

As we can see from Figure~\ref{fig:specfit}(f), the $\chi^2$ of the single-cloud fitting is 791.67, equivalent to a reduced $\chi^2$ of 7.01 since the degree-of-freedom is 113. Nevertheless, the reduced $\chi^2$ of the two-cloud fitting is 0.57 with a degree-of-freedom of 109, meaning the two-cloud model more appropriate. We then fit the H$\alpha$ profiles of the studied filament, as shown typically in Figures~\ref{fig:specfit}(d)--(f), using Equations (\ref{eq:one_fil}) and (\ref{eq:two_fil}). The averaged fitting uncertainty of the Doppler velocity is 0.66 km s$^{-1}$, combining the uncertainty of wavelength calibration of 1.12 km s$^{-1}$, leading to the total uncertainty 1.30 km s$^{-1}$. It is not surprising that the fitting error of the filament is much smaller than the prominence because the spectral profile of filament is much smoother and less influenced by the noise on the continuum part.

\begin{figure}[ht!]
	\plotone{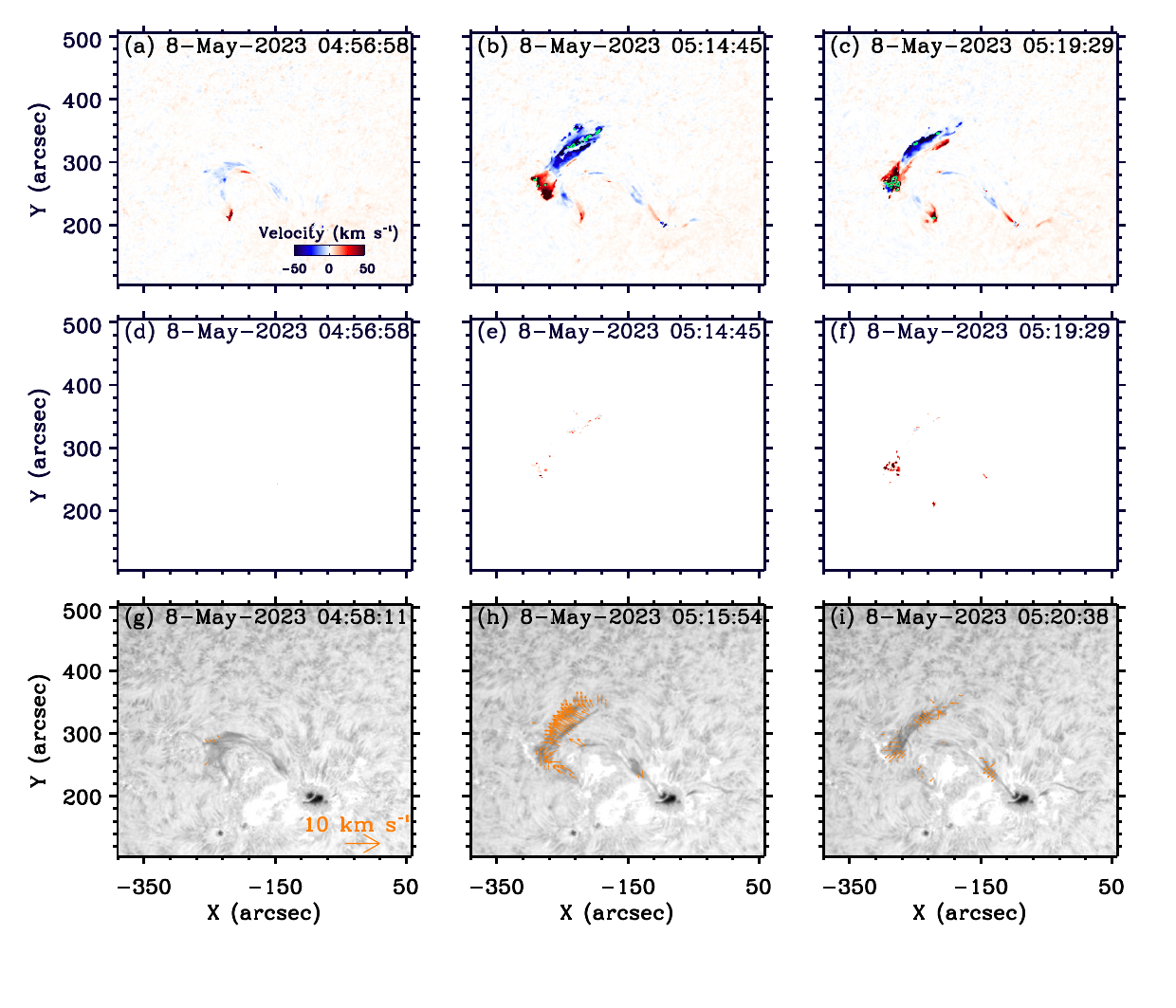}
	\caption{Three-dimensional velocity fields of the filament. (a)--(c) Doppler velocity of both the single-cloud model and the front cloud in the double-cloud model. The green lines show the region using double-cloud model. (d)--(f) Doppler velocity of the back cloud in the double-cloud model. (g)--(i) POS velocity with the orange arrows indicating the directions. A 2-second animation is available online, showing the evolution of the filament's three-dimensional velocities from 04:58:11 UT to 05:25:22 UT.	\label{fig:filament2}}
\end{figure}

Figures~\ref{fig:filament2}(a)--(f) show the Doppler velocity fields of the ejecting filament (corresponding animation is available online). It seems that this filament splits discretely during the propagation. In the early stage, the filament shows a global blueshift with a growing Doppler velocity from -4 km s$^{-1}$ to -70 km s$^{-1}$. At 05:07:39 UT, the southeast leg turns to be redshifted. Then the top of the filament consists of two parts with blueshift on the front side and redshift on the other side since 05:10:01 UT. Until 05:24:13 UT when all the material starts to fall back, the filament appears as a holistic redshift, resulting in a failed eruption.


\subsection{POS velocity of the eruptive filament}
The POS velocity of the filament is computed in the same way as the prominence in section~\ref{subsec:transverse_pro}. The results are displayed in Figures~\ref{fig:filament2}(g)--(i). At first, the filament expands outward as it ascends. Then at 05:11:10 UT, the arrows at the north west begin to point to the inside, opposite to the expansion direction. The inward POS velocity, along with the Doppler pattern which shows a blueshifted front and a redshifted back, reveals that the filament appears to rotate counterclockwise viewed from the main positive polarity, also indicating that it is a sinistral filament as mentioned in section~\ref{sec:data}. Combining the LOS and POS velocities, the maximum ejecting velocity of the filament is estimated to be approximately 71 km s$^{-1}$ with 81$^\circ$ inclined to the POS, and it begins to drop back to the Sun in about 16 minutes after the eruption onset.


\section{Discussion and Conclusion}\label{sec:dis_con}
In this Letter, we employ the full-disk high-resolution spectra detected by CHASE to reconstruct three-dimensional velocity fields of a prominence eruption on 2022 August 17 and a failed filament eruption on 2023 May 8, respectively. The cloud model is used to derive the LOS velocities, and the FLCT method is used to calculate the POS velocities.  

The derived three-dimensional kinematics of the prominence show that, in the early phase of the eruption, the prominence expands outward in a shape close to semicircle. Then, some parts of the prominence split into two pieces with different Doppler velocities along the LOS direction, resulting in the bimodal H$\alpha$ line profiles. As the ejection goes on, the splitting gets more violent and extends from the top of the prominence to almost the whole prominence. A similar phenomenon had been observed by the Interface Region Imaging Spectrograph (IRIS) using cool lines at chromopheric and transition region with a slit-stare mode \citep{2015ApJ...803...85L}. The two-peak profiles show up at the beginning and the two pieces separate from each other gradually, so the authors inferred that that prominence is a hollow cone structure.  \cite{2018ApJ...865..123R} also reported that the bimodal spectral profiles might exist in the quiescent prominence because of the prominence fine structures situating variously along the LOS. However, the prominence studied here reveals that the splitting occurs during the eruption and evolves with time.In the later phase, the south leg of the prominence moves downward in a swirling way because of the gravity. The maximum velocity of the prominence is approximately 160 km s$^{-1}$ with an inclination angle of 22$^\circ$ to the POS and the propagation velocity of the prominence top keeps growing until turning into a CME. 

The lifting filament on 2023 May 8 expands outward in a semicircle shape at the first time. The expansion velocity is 71 km s$^{-1}$ with an inclination angle of 81$^\circ$ to the POS, almost along the LOS direction. The speed is much smaller than the prominence eruption on 2022 August 17. Moreover, the filament falls back to the solar surface in a short time, thus no CMEs is produced. As indicated by the two-cloud fitting result, there are sporadic material splitting along the LOS direction during the eruption, various from the blocky splitting of the prominence. In addition, the draining sites are right-skewed according to the conjugate magnetic polarity distribution. The three-dimensional velocities meanwhile illustrate the counterclockwise rotation of the material in the filament. Both phenomena indicate that this filament is of sinistral chirality, deviating from the hemispheric rule \citep{2017ApJ...835...94O}. 

There had been several three-dimensional kinematic reconstructions done before. \cite{2003PASJ...55..503M}, \cite{2017ApJ...836...33C} and \cite{2021PASJ...73..394G} adopted the single-cloud model and Local Correlation Tracking (LCT) method to the FMT data. The FMT H$\alpha$ observation has only three wavelength points, but there are four free parameters in the single-cloud model. Thus, some parameters were fixed by the data before the eruption in their computations. \cite{2021EP&S...73...58S} employed the SDDI data with the wavelength sampling of 0.25~\AA~by tracking the position and Doppler velocity at the filament apex. The low spectral resolution of such instruments leads to large uncertainties of the calculated LOS velocities. Besides the H$\alpha$ waveband instruments, \cite{2015ApJ...803...85L}, \cite{2017A&A...606A..30S}, \cite{2019A&A...624A..72Z} and \cite{2023Ap&SS.368...40J} used the IRIS chromspheric spectral lines to obtain the velocities of the active prominences. The spatial and spectral resolutions of IRIS are 0.33$''$-0.4$''$ and 0.026-0.053~\AA. The FOV is only 175$''$ $\times$ 175$''$ \citep{2014SoPh..289.2733D}, unable to cover the whole filament. Therefore, the CHASE full-disk high-resolution spectra is valuable in the application of computing three-dimensional velocity fields of filaments. With the derived three-dimensional kinematics of filament eruption, we can further investigate its triggering mechanism, the causes of material splitting, the MHD instabilities inside the filament, and its relationship with the CMEs.

\begin{acknowledgments}
The CHASE mission is supported by China National Space Administration. SDO is a mission of NASA's Living With a Star Program. This study is supported by NSFC (12333009, 12233012, and 12127901) and the National Key R\&D Program of China (2022YFF0503004, 2021YFA1600504, and 2020YFC2201200), and CNSA project D050101.

\end{acknowledgments}


\bibliography{manuscript}{}
\bibliographystyle{aasjournal}

\end{document}